%% file: main.tex
\documentclass[12pt]{article}
\title{}
\author{}
\date{}
\usepackage[utf8]{inputenc}
\usepackage[letterpaper,top=1in,right=1in,left=1in,bottom=1in]{geometry}
\usepackage{amsmath}
\usepackage{amsfonts}
\usepackage{amssymb}
\usepackage{graphics}
\usepackage[demo]{graphicx}
\usepackage{placeins}
\usepackage[capposition=top]{floatrow}
\usepackage{subcaption}
\usepackage{lscape}
\usepackage{pdflscape}
\usepackage{setspace}
\usepackage{standalone}
\usepackage{fancyhdr}
\usepackage{framed}
\usepackage{color}
\usepackage{tikz}
\usepackage{pgfplots}
\usepackage{longtable}
\usepackage{booktabs,caption}
\usepackage[draft]{todonotes}
\usepackage{booktabs,caption}
\usepackage[flushleft]{threeparttable}
\usepackage{bbm}
\usepackage{multirow}
\usepackage{lineno}
\usepackage{titlesec}
\usepackage{hyperref}
\usepackage{comment}
\usepackage[draft]{todonotes}
\usepackage[justification=centering]{caption}
\pgfplotsset{compat=1.15}

\usepackage{longtable}
\usepackage{array}
\usepackage{authblk}
\usepackage{pdflscape}
\usepackage{colortbl}
\usepackage{ragged2e}
\usepackage{xfp}

\usepackage{stackengine}
\usepackage{array}
\newcolumntype{L}[1]{>{\raggedright\arraybackslash}p{#1}}
\setstackEOL{\#}
\setstackgap{L}{12pt}

\usepackage[page,toc,titletoc,title]{appendix}
\usepackage{blindtext}
\usepackage{pdfpages}

\hypersetup{
	colorlinks=true, 
	linktoc=all,     
	linkcolor=black,  
	citecolor=black,
    urlcolor=blue
}

\usepackage[style=apa,natbib]{biblatex}
\usepackage[all]{nowidow} 
\usepackage[protrusion=true,expansion=true]{microtype} 

\usepackage{orcidlink}

\hypersetup{ 	
pdfsubject = {},
pdftitle = {SSRI HTE},
pdfauthor = {Gilbert, Joshua B.}
}

\usepackage{setspace}

\title{Item-Level Heterogeneous Treatment Effects of Selective Serotonin Reuptake Inhibitors (SSRIs) on Depression: Implications for Inference, Generalizability, and Identification}

\begin{document}

\begin{titlepage}
\author[1]{Joshua B. Gilbert\,\orcidlink{0000-0003-3496-2710}}
\author[2]{Fredrik Hieronymus\, \orcidlink{0000-0003-0930-6068}}
\author[2]{Elias Eriksson}
\author[3]{Benjamin W. Domingue\,\orcidlink{0000-0002-3894-9049}}
\affil[1]{Harvard Graduate School of Education}
\affil[2]{University of Gothenburg}
\affil[3]{Stanford Graduate School of Education}

\maketitle

\begin{abstract}

\noindent \textbf{Objectives:} In analysis of randomized controlled trials (RCTs) with patient-reported outcome measures (PROMs), Item Response Theory (IRT) models that allow for heterogeneity in the treatment effect at the item level merit consideration. These models for ``item-level heterogeneous treatment effects'' (IL-HTE) can provide more accurate statistical inference, allow researchers to better generalize their results, and resolve critical identification problems in the estimation of interaction effects. In this study, we extend the IL-HTE model to polytomous data and apply the model to determine how the effect of selective serotonin reuptake inhibitors (SSRIs) on depression varies across the items on a depression rating scale. \\

\noindent \textbf{Methods:} We first conduct a Monte Carlo simulation study to assess the performance of the polytomous IL-HTE model under a range of conditions. We then apply the IL-HTE model to item-level data from 24 RCTs measuring the effect of SSRIs on depression using the 17-item Hamilton Depression Rating Scale (HDRS-17) and estimate heterogeneity by subscale (HDRS-6). \\

\noindent \textbf{Results:} Our simulation results show that ignoring IL-HTE can yield standard errors that are as much as 50\% too small and create significant bias in treatment by covariate interaction effects when item-specific treatment effects are correlated with item location, and that the application of the IL-HTE model resolves these issues. Our empirical application shows that while the average effect of SSRIs on depression is beneficial (i.e., negative) and statistically significant, there is substantial IL-HTE, with estimates of the standard deviation of item-level effects nearly as large as the average effect. We show that this substantial IL-HTE is driven primarily by systematically larger effects on the HDRS-6 subscale items. \\

\noindent \textbf{Conclusions:} The IL-HTE model has the potential to provide new insights for the inference, generalizability, and identification of treatment effects in clinical trials using PROMs. \\


\centering
Forthcoming in \textit{Epidemiologic Methods}
\end{abstract}

\end{titlepage}

\doublespacing

\section{Introduction}

Heterogeneous treatment effects (HTE) are crucial for epidemiological research and public health policy because understanding for whom and under what conditions a medical treatment works allows policy makers to best target treatments to populations or subgroups that would benefit the most \citep{beghi2011epidemiology, cordero2021key, kent2016risk, varadhan2013framework, robertson2021assessing, lesko2018considerations}. One limitation of standard statistical methods for HTE analysis is that they focus on person characteristics (e.g., age, gender, etc.) and may ignore the potential HTE that exists among the items used to measure a latent variable of interest. That is, many outcomes relevant to clinical trials or epidemiological research can only be assessed indirectly through multi-item surveys or psychometric instruments, such as patient-reported outcome measures (PROMs) for well-being \citep{mcevoy2011epidemiology}, depression \citep{hieronymus2019influence, sajobi2023unsupervised}, perceptions of hearing loss \citep{jessen2018improving}, pain \citep{dworkin2009development}, health literacy \citep{woods2021cluster}, or recovery after childbirth \citep{sultan2020evaluation, sultan2021use}, in contrast to simple biometric measures that can be measured to arbitrary levels of precision and captured in a single number (e.g., height, weight, blood pressure, mortality, etc.). Therefore, when we use a PROM to construct a sum or factor score to serve as an outcome measure in clinical trials or epidemiological studies, we may be ignoring the potential HTE that exists among the individual items of the PROM. As a result, our understanding of the consistency or generalizability of treatment effects may be limited \citep{gilbert2023modeling, ahmed2023heterogeneity, sales2021effect}, and---when there is interest in treatment by baseline covariate interactions---creates challenges to causal identification that can only be resolved by leveraging item-level data \citep{gilbert2023disentangling}.

To address these limitations, recent work applies techniques from Item Response Theory (IRT; \cite{van2017handbook1}) to allow for the assessment of treatment effects that vary at the outcome item level \citep{sales2021effect, ahmed2023heterogeneity, gilbert2023modeling}. These techniques allow us to determine whether the treatment effects are consistent and impact all PROM items equally or, alternatively, vary across the items within the PROM. However, to our knowledge, the ``item-level heterogeneous treatment effects'' (IL-HTE) model has not yet been applied beyond standardized tests in education research; clinical trial or epidemiological studies that similarly rely on latent variable outcomes such as PROMs may also benefit from such approaches. Our work therefore builds on previous interest in the affordances of item-level analysis in clinical trials and epidemiological research \citep{hieronymus2019influence, jones2019differential, grayson2000item, barger2023epidemiology, chan2004interview, andresen2013performance}.

Our study pursues two aims. First, we apply the IL-HTE model to polytomous item response data, thereby extending past analyses limited to dichotomous (i.e., correct vs. incorrect) item responses. Polytomous data is of interest in epidemiological settings given the widespread use of, for example, Likert scales \citep{capuano2016modeling, esterman2003likert}. Second, we apply the IL-HTE model to a clinical trial context using item-level data from a set of 24 randomized controlled trials (RCTs) evaluating the effect of SSRIs on the 17-item Hamilton Depression Rating Scale (HDRS-17). The study is organized as follows. In Section \ref{hte}, we contextualize the IL-HTE model within the broader context of HTE analysis, for both dichotomous and polytomous responses. In Section \ref{methods}, we describe our Monte Carlo simulation design and our empirical data. In Section \ref{results}, we present the results of the simulation and the results of our models fit to the empirical data. We conclude in Section \ref{discussion} with a discussion of the implications of our findings for the analysis of HTE in clinical trials and epidemiology.

\subsection{Estimating Heterogeneous Treatment Effects}\label{hte}

Consider the following standard regression model for HTE:

\begin{align}
\label{eq_hte}
    Y_j &= \beta_0 + \beta_1 T_j + \beta_2 X_j + \beta_3 T_j \times X_j + \varepsilon_j \\
    \varepsilon_j &\sim N(0,\sigma_\theta),
\end{align}

\noindent in which $Y_j$ is the outcome variable for individual $j$, $T_j$ is a dichotomous indicator for randomized treatment assignment, and $X_j$ is a person characteristic (e.g., age, gender, etc.). $\beta_1$ is the conditional average treatment effect (CATE) when $X=0$, $\beta_2$ is the main effect of $X$ when $T=0$, and $\beta_3$ is our HTE parameter, capturing how the CATE depends on the level of $X$. When $\beta_3=0$, the ATE is constant across all level of $X$; when $\beta_3 > 0$, the ATE is larger at higher levels of $X$. A concrete example of such a model in epidemiology would be how the effect of COVID vaccinations ($T$) on mortality ($Y$) varies by patient age ($X$) \citep{collier2021age, faro2022population}.

When $Y_j$ is a latent outcome such as a PROM, such as a depression rating scale with multiple items reflecting various symptoms of depression, treatment may differentially affect individual symptoms or clusters of symptoms represented by the individual items of the PROM. The standard HTE model above represents one extreme based on analysis of a single number such as the sum score as the outcome (a common practice; \cite{flake2017construct, mcneish2020thinking}) that ignores all such differentiation. On the other extreme, researchers could analyze treatment effects on each item separately, but this approach is difficult to interpret and suffers from multiple comparisons problems, particularly when the number of items is large. As a compromise, researchers examine effects on subscales or item clusters, but this approach requires an \textit{a priori} specification of which subscales to evaluate and the assumption that within each subscale, the item effects are constant \citep{gilbert2023tutorial}, and as such is somewhat ad hoc.

An elegant solution that leverages the PROM item responses directly and thus makes use of all available information without the need to compute total or subscale scores in a separate step of the analysis is the explanatory item response model (EIRM) \citep{wilson2008explanatory, wilson2004descriptive, petscher2020past, de2016explanatory}. For example, when items are dichotomous (e.g., 0 = symptom absence, 1 = symptom presence), we can use a cross-classified logistic regression model with a main effect for treatment, such as,
\begin{align}
\label{constant_eq}
    \text{logit}( P(Y_{ij}=1)) = \eta_{ij} &= \theta_j + b_i \\
    \theta_j &= \beta_0 + \beta_1 T_j + e_j \\
    b_i &= b_{i} \\
    b_{i} &\sim N(0,\sigma_b) \\
    e_j &\sim N(0,\sigma_\theta),
\end{align}

\noindent where $Y_{ij}$ is the response of person $j$ to item $i$, $\theta_j$ is the unobserved or latent person trait (e.g., depression), and $b_i$ is item location (i.e., ``easiness'' in educational measurement). The explanatory item response model is equivalent to a one-parameter logistic (1PL) or Rasch IRT model when the item location parameters are considered fixed \citep{de2008random}. $\beta_1$ represents the ATE, but estimated directly on the latent trait $\theta_j$ without the need to compute a sum or factor score to be used as an outcome in a two-step analysis \citep{gilbert2023tutorial, gilbert2023modeling, gilbert2024estimating, gilbert2023measurement, christensen2006rasch, zwinderman1991generalized}. Even without considering the possibility of IL-HTE, the explanatory item response model still provides some benefits over a sum score analysis as it can be more robust to violations of model assumptions such as heteroskedasticity or missing data \citep{gilbert2024estimating} and can provide unbiased estimates of standardized effect sizes that are attenuated by measurement error \citep{gilbert2023measurement, hedges1981distribution}.

We can allow for IL-HTE by introducing a random interaction between treatment and item in a random slope term in the equation for $b_{ij}$, where the added subscript $j$ allows for separate $b$ parameters for each treatment group \citep{gilbert2023modeling,gilbert2023tutorial, ahmed2023heterogeneity, sales2021effect, gilbert2023disentangling}:

\begin{align}
\label{il_hte_eq}
    \text{logit}( P(Y_{ij}=1)) = \eta_{ij} &= \theta_j + b_{ij} \\
    \theta_j &= \beta_0 + \beta_1 T_j + e_j \\
    \label{b_eq}
    b_{ij} &= b_i + \zeta_{i} T_j \\ 
    \begin{bmatrix}
        b_i \\
        \zeta_{i}
    \end{bmatrix}
     &\sim N(0,\begin{bmatrix}
         \sigma_b & \rho \\
         \rho & \sigma_\zeta
     \end{bmatrix}) \\
    e_j &\sim N(0,\sigma_\theta).
\end{align}

\begin{figure}
    \centering
    \includegraphics[width=1\linewidth]{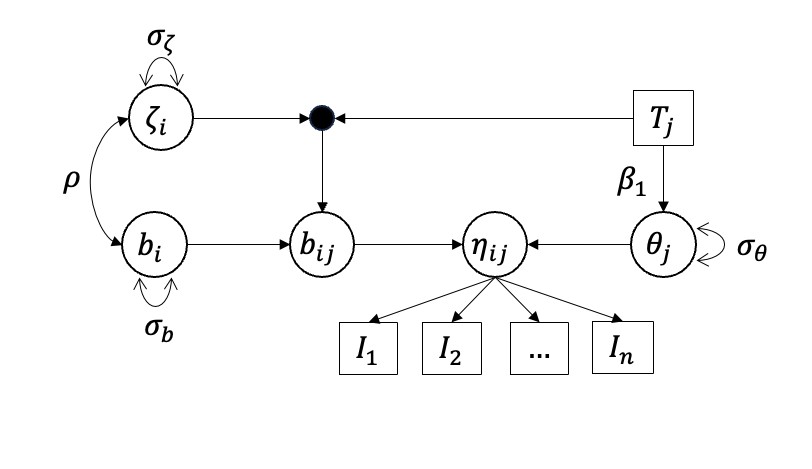}
    \caption{Directed Acyclic Graph of the IL-HTE Model}
        \justify \footnotesize 
    Notes: Squares indicate observed variables, hollow circles indicate latent variables, and solid circles represent cross product interaction terms. $I_n$ are item responses and $T_j$ is the treatment indicator. $\beta_1$ represents the average treatment effect. $\rho$ represents the correlation between item location and item-specific treatment effect size. Path coefficients are fixed at 1 unless otherwise indicated.
    \label{fig:dag}
\end{figure}

\noindent Here, $\beta_1$ still represents the ATE on the latent trait $\theta_j$, but now for the \textit{average} item on the scale. Item-specific residual treatment effects are represented by $\zeta_i$, the item-specific deviation from $\beta_1$. If $\zeta_i > 0$, then item $i$ is more affected by treatment than the average item. The $\zeta_i$ are equivalent to uniform differential item function (DIF) caused by the treatment \citep{gilbert2023modeling, montoya2020mimic}, and $\beta_1 + \zeta_i$ represents the total treatment effect size for item $i$. $\sigma_\zeta$ provides a direct parameter estimate for the degree of IL-HTE in the data by providing the standard deviation (SD) of the item-specific treatment effects around the ATE $\beta_1$ and $\rho$ indexes the correlation between item location and item-specific treatment effect size. That is, if $\rho>0$, then items representing more commonly endorsed symptoms show systematically larger or smaller treatment effects. $\rho$ may be of interest in itself \citep[pp. 893-894]{gilbert2023modeling}, but is perhaps most critical in that $\rho \neq 0$ can create a causal identification problem. That is, $\rho \neq 0$ can induce spurious treatment by baseline covariate interaction effects that can only be resolved with item-level analysis \citep{gilbert2023disentangling}, a result with critical implications for the analysis of HTE that we will return to in our simulations, particularly given the great interest in HTE by baseline severity in depression treatments \citep{kirsch2008initial, fournier2010antidepressant}. A directed acyclic graph (DAG) \citep{greenland1999causal, joffe2012causal, glymour2006using, tennant2021use} representation of the IL-HTE model is presented in Figure \ref{fig:dag}.

A large value of $\sigma_\zeta$ suggests that there is substantial IL-HTE in the data. We can test whether item features can explain the IL-HTE by interacting item properties (e.g., item subscale, item type, item modality, etc.) with the treatment indicator to explain some of the IL-HTE. For example, we can extend Eqn. \ref{b_eq} as follows:

\begin{align}
\label{subscale_eq}
    \theta_j &= \beta_0 + \beta_1 T_j + e_j \\
    b_{ij} &= b_i + \gamma_1 S_i + \gamma_2 S_i \times T_j + \zeta_{i} T_j.
\end{align}

\noindent Here, $S_i$ is an indicator for whether item $i$ is part of subscale $S$. Accordingly, $\beta_1$ represents the CATE for items in the reference set (assumed here to be impacting $\theta_j$ directly), $\gamma_1$ is a main effect for item location, and $\gamma_2$ provides the difference in the CATE for items on subscale $S$. For example, education researchers have used the subscale effects IL-HTE model to examine treatment effects on items related to different reading comprehension passages \citep{gilbert2023modeling, gilbert2023tutorial, kim2023longitudinal}. We can use such subscale models to calculate the proportion of IL-HTE variance explained by the treatment by item characteristic interaction term $S_i \times T_j$ with a pseudo-$R^2$ metric by comparing the reduction in $\sigma^2_\zeta$ from the unconditional IL-HTE model (Eqn. \ref{il_hte_eq}) to the subscale interaction model (Eqn. \ref{subscale_eq}) using the following equation:

\begin{align} \label{eq_r2}
    R^2 = \frac{\sigma^2_{\zeta_0} - \sigma^2_{\zeta_1}}{\sigma^2_{\zeta_0}},
\end{align}

\noindent where $\sigma^2_{\zeta_0}$ is derived from the unconditional IL-HTE model and $\sigma^2_{\zeta_1}$ is derived from the subscale interaction model.

The application of the explanatory item response model to polytomous responses is less common, but it is straightforward to extend the IL-HTE model to polytomous data. In short, we can leverage the computational machinery of binary logistic regression to fit polytomous models by reshaping the data to represent pairwise contrasts between ordered response categories. We then fit a logistic regression model that includes fixed effects for item threshold parameters that represent the boundaries between each cut point on the scale \citep{bulut2021estimating}. If we assume that the distances between thresholds are equal across items, the result is a Rating Scale Model (RSM). In contrast, if we assume that each item has a unique distance between each threshold, the result is a Partial Credit Model (PCM). While such models are typically implemented with fixed item and threshold parameters, they can be extended to the random item case that better allows for IL-HTE modeling \citep{kim2020polytomous}. For readers interested in applying these models to their own data sets, our references provide various tutorials in the R programming language: the general explanatory item response model for dichotomous items \citep{de2011estimation}, extending the explanatory item response model to polytomous items \citep{bulut2021estimating}, the IL-HTE model for dichotomous items \citep{gilbert2023tutorial}, and a Bayesian approach that allows for extensions such as 2PL models or the Graded Response Model (GRM) \citep{burkner2021brms, gilbert2023tutorial}.

\section{Methods}\label{methods}

\subsection{Monte Carlo Simulation}

We use Monte Carlo simulation to examine the performance of the IL-HTE model applied to polytomous item responses. Previous simulation studies of the IL-HTE model using dichotomous item responses have demonstrated two key results. First, in terms of statistical inference, they show that substantial IL-HTE increases the standard error of the ATE ($\text{SE}(\hat\beta_1)$) but does not cause bias \citep{gilbert2023modeling}. Second, in terms of causal identification, they show that the correlation $\rho$ between item location $b_i$ and item-specific treatment effect size $\zeta_i$ induces spurious treatment by covariate interaction effects, which is critical because it suggests that standard HTE analysis of composite outcomes (e.g., Eqn. \ref{eq_hte}) can be misleading \citep{gilbert2023disentangling}. We hypothesize that the same pattern of results will apply to polytomous data, because ordered logit models are approximately invariant to the collapsing of response categories \citep[p. 13]{steele2011module}.

We simulate polytomous item responses from the Rating Scale Model and fit each model with the \texttt{glmer} function from the \texttt{lme4} package in R with fixed item thresholds and random item location parameters \citep{gilbert2023tutorial, bulut2021estimating, de2011estimation, kim2020polytomous}. We fit two sets of simulations, one to examine the effect of IL-HTE on ATEs and associated SEs (i.e., replicating and extending \cite{gilbert2023modeling}), and the other to examine the effect of $\rho $ on treatment by covariate interaction effects (i.e., replicating and extending \cite{gilbert2023disentangling}). We fix the following parameters in all our simulations: an ATE of .20 on the logit scale, $\sigma_b=1$, $\sigma_\theta = .5$, and a baseline covariate with a coefficient of 1. 

In our first set of simulations, we vary the following factors in a fully crossed design: the number of categories $k$ at 3, 5, 7, $\sigma_\zeta$ at 0, .2, .4 SDs representing no, moderate, and large IL-HTE, sample sizes of 300, 500, and 1000 patients to represent small, moderate, and large samples, and PROM lengths of 8, 12, and 20 items to represent short, moderate, and long assessments. We fit two models to each simulated data set, one assuming a constant treatment effect, the other allowing for IL-HTE. In our second set of simulations, we fix $k=3$ and $\sigma_\zeta=.4$ and vary $\rho$ from -1 to 1 in increments of .25 to determine how an estimated treatment by baseline covariate interaction becomes biased when we assume a constant treatment effect model. We fit two models to each simulated data set, one allowing for a treatment by baseline covariate interaction but constant item effects, the other allowing for both the interaction and IL-HTE. We repeat the process 200 times per condition.

\subsection{Empirical Application}

\subsubsection{The Hamilton Depression Rating Scale (HDRS)}

The Hamilton Depression Rating Scale (HDRS) has been the de facto``gold standard'' in antidepressant research for more than half a century \citep{ruhe2005clinical}. The HDRS has its origins in the 1950s in England and has since been applied worldwide \citep{hamilton1960, obeid2018validation}. The most widely used version of the HDRS, used here, consists of 17 polytomous (3- and 5-category) items \citep{williams2001} (HDRS-17). 

There is a long history of psychometric analysis of the HDRS-17 and derivative measures, including questioning its psychometric properties \citep{bagby2004} or focusing on dimensionality and applying various techniques aimed at distinguishing an HDRS subscale that can function as a unidimensional measure of the severity of depression between different populations and treatments \citep{bech1975, gibbons1993exactly,bech1981hamilton}. Prior analyses of item-level HDRS data include the reliability of subscales or items \citep{luckenbaugh2015rating}, treatment heterogeneity by baseline depression severity \citep{hieronymus2019influence}, the properties of the unidimensional HDRS-6 subscale (also known as the Bech or melancholia subscale; items 1, 2, 7, 8, 10, 13) \citep{bech1981hamilton, rush2021clinically, park2017clinical}, and patterns of treatment effects when individual items or subscales are analyzed separately \citep{hieronymus2015}. We emphasize that our analysis of the HDRS-17 data is intended to be illustrative of the affordances of the IL-HTE model rather than a definitive analysis of the measurement properties of the HDRS-17 scale. The HDRS-17 items are summarized in Table \ref{tab:hdrs17}.

\begin{table}
    \centering
    \begin{tabular}{cccc}
         Item Number&  Subject & HDRS-6 Item &Range\\
         \hline
         1&  Depressed Mood & Yes &0-4\\
         2&  Feelings of Guilt & Yes &0-4\\
         3&  Suicide & &0-4\\
         4&  Insomnia: Early Night & &0-2\\
         5&  Insomnia: Middle Night & &0-2\\
         6&  Insomnia: Early Morning & &0-2\\
         7&  Work and Activities & Yes &0-4\\
         8&  Retardation & Yes &0-4\\
         9&   Agitation& &0-4\\
 10&  Anxiety: Psychic&Yes &0-4\\
 11&  Anxiety: Somatic&&0-4\\
 12&  Gastro-Intestinal Symptoms&&0-2\\
 13&  General Somatic Symptoms&Yes &0-2\\
 14&  Genital Symptoms&&0-2\\
 15&  Hypochondriasis&&0-4\\
 16&  Loss of Weight&&0-2\\
 17&  Insight&&0-2\\
 \hline
    \end{tabular}
    \caption{HDRS-17 Items}
    \label{tab:hdrs17}
\end{table}

\subsubsection{Data}

For our empirical application, we use a subset of data from a previous study that evaluated the effects of SSRIs on depression measured with the HDRS-17 \citep{hieronymus2019influence}. Our sample consists of data from 24 RCTs that compare acute phase SSRI treatment with placebo in patients diagnosed with depression. 8262 participants were included in the studies and in this analysis we focus on those 5313 patients who had HDRS-17 data available after six weeks of treatment (i.e., excluding patients who dropped out of treatment due to, for example, adverse events or lack of efficacy; for further details see \cite{hieronymus2015, hieronymus2019influence}). In total, 90313 person-item combinations were included. There were 8 missing item responses; one advantage of the explanatory item response model approach is that it employs Maximum Likelihood for missing item response data \citep{gilbert2024estimating}. We included baseline HDRS-17 sum scores as a covariate to improve the precision of our estimates.

\subsubsection{Model Building Strategy}

We fit a taxonomy of five models, summarized in Table \ref{tab:mods}. Models 1A and 1B use the standardized HDRS-17 sum score as an outcome variable in a standard regression framework. Model 1A assumes a constant treatment effect across baseline depression, and Model 1B allows for a treatment by baseline depression interaction, as in Equation \ref{eq_hte}. We begin with the sum score approach to illustrate the standard practice and provide a benchmark for comparison. Models 2A, 2B, and 2C are Rating Scale Models that model the polytomous item response data directly, using fixed effects for average item thresholds with random uniform item location shifts. Model 2A allows for a constant treatment effect (Eqn. \ref{constant_eq}), Model 2B allows for randomly varying IL-HTE by including a random slope for treatment at the item level (Eqn. \ref{il_hte_eq}) and Model 2C adds a main effect and interaction for the HDRS-6 subscale items (Eqn. \ref{subscale_eq}) to determine whether treatment effects vary systematically by subscale.

\begin{table}
    \centering
    \begin{tabular}{ccl}
         Label& Model&Treatment Effect\\
         \hline
         1A& Sum Score&Constant TE\\
         1B& Sum Score&Treatment by baseline interaction\\
         2A& RSM&Constant TE\\
         2B& RSM&Randomly Varying IL-HTE\\
 2C&RSM&Subscale Effects IL-HTE\\
 \hline
    \end{tabular}
    \caption{Taxonomy of regression models fit to the SSRI Data}
            \justify \footnotesize 
    Notes: RSM = Rating Scale Model.
    \label{tab:mods}
\end{table}

We also examined two additional modeling strategies to probe the sensitivity of our results. First, to examine how $\rho$ can induce spurious interaction effects, we fit an additional set of models that interact treatment status with baseline depression to determine how sensitive the model is to the inclusion or exclusion of $\rho$. Second, while the Rating Scale Model is not typically applied when items have different numbers of response categories, we determined that it would be appropriate in our case because item responses with ratings of 4 or 5 were (a) quite rare in our sample (see online supplemental materials) and (b) represented a qualitatively more extreme range of symptoms than the 3-category items. In our online supplement, we fit the more flexible Partial Credit Model that allows for separate thresholds for each item and show that the results of our analysis are unchanged. 

\section{Results}\label{results}

\subsection{Monte Carlo Simulation}

\begin{figure}[tb]
    \centering
    \includegraphics[width=1\linewidth]{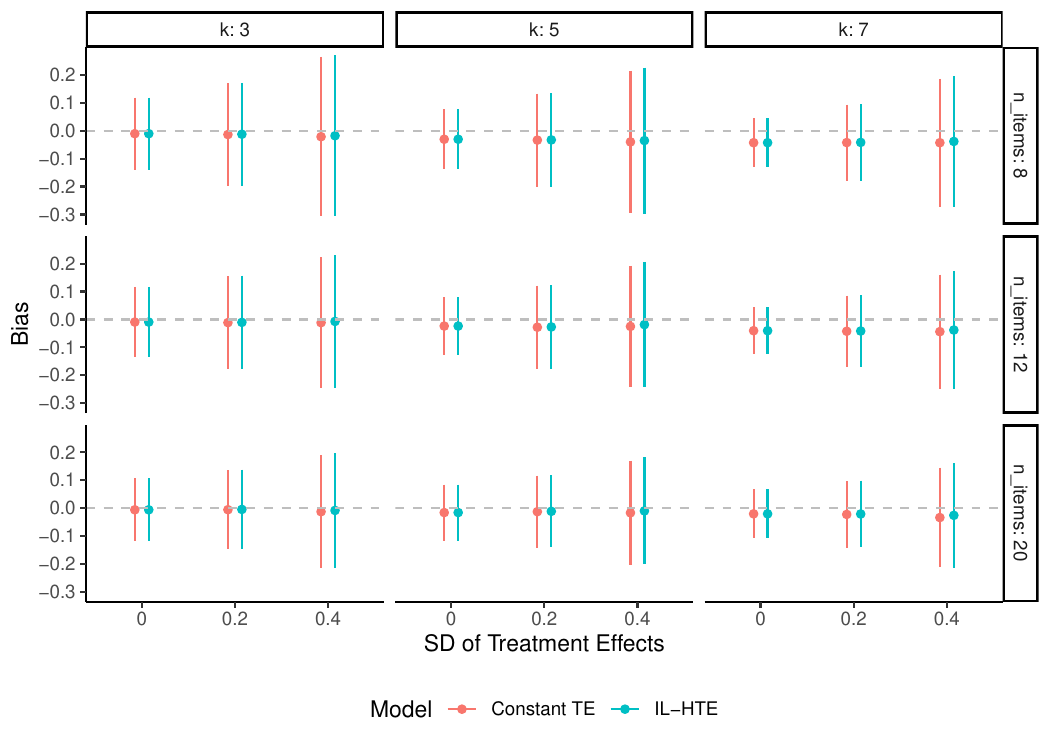}
    \caption{Estimated Bias of Treatment Main Effect by Simulation Condition}
               \justify \footnotesize 
    Notes: The y-axis shows average treatment effect bias ($\hat\beta_1 - \beta_1$) and the x-axis shows the SD of the item-specific treatment effects ($\sigma_\zeta$). The points and bars represent the mean and $\pm$2 SDs, respectively. \textit{k} is the number of item response categories and \textit{n\_items} is the number of items. The results are averaged across the sample size conditions.
    \label{fig:bias_box}
\end{figure}

\begin{figure}[tb]
    \centering
    \includegraphics[width=1\linewidth]{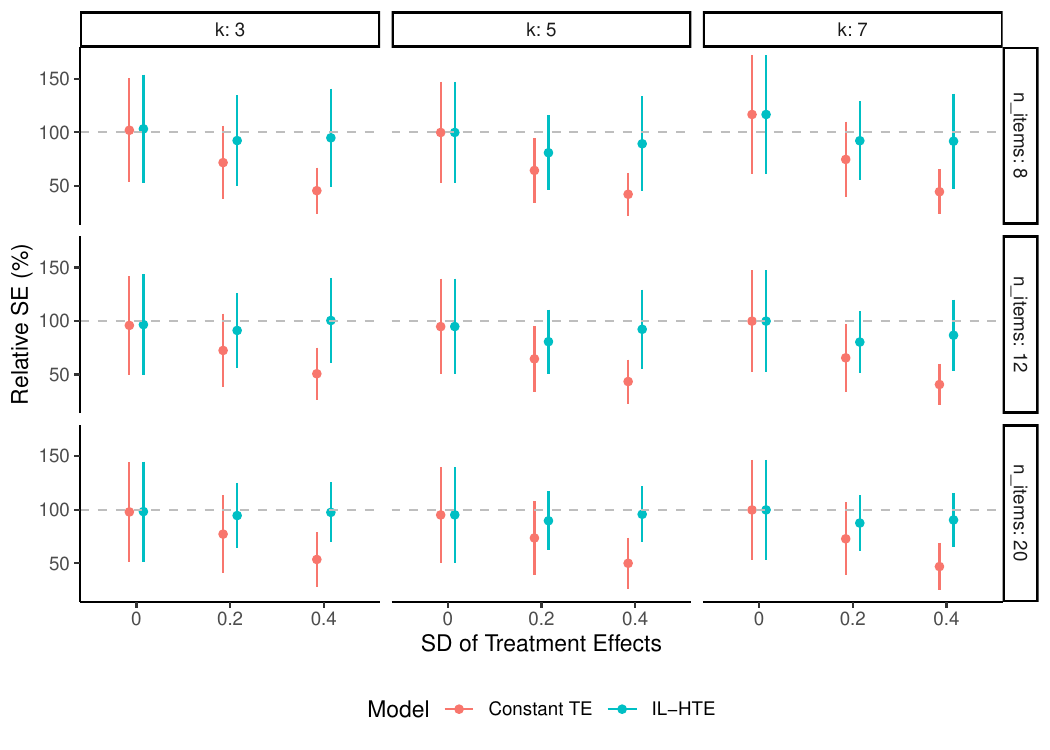}
    \caption{Estimated Standard Error Calibration of Treatment Main Effect by Simulation Condition}
        \justify \footnotesize 
    Notes: The y-axis shows the relative standard error, such that 100\% indicates that the model-based SE was equivalent to the empirical SE, on average, and the x-axis shows the SD of the item-specific treatment effects ($\sigma_\zeta$). The points and bars represent the mean and $\pm$ 2 SDs, respectively. \textit{k} is the number of item response categories and \textit{n\_items} is the number of items. The results are averaged across the sample size conditions.
    \label{fig:se_cal}
\end{figure}

Our simulation results replicate previous findings from the dichotomous IL-HTE model in the polytomous setting \citep{gilbert2023modeling}. That is, in terms of statistical inference, IL-HTE does not cause substantial bias in the treatment effect point estimates, but IL-HTE vastly inflates the SE of the ATE $\beta_1$. Figure \ref{fig:bias_box} shows the distribution of bias by condition and we see that the degree of bias is small across all conditions and does not vary by model type. While there is a slight negative bias in both models and at all levels of IL-HTE as the number of response categories increases, the absolute magnitude of the bias is small, and additional simulations in our online supplement show that the magnitude of the bias approaches 0 as the number of items increases. Figure \ref{fig:se_cal} shows the calibration of the model SEs by plotting the mean model-based SE against the observed SD of the treatment effect point estimates (i.e., the empirical SE). A ratio of 100\% indicates that the mean model-based SE is equivalent to the empirical SE. We clearly see that as IL-HTE increases, the constant TE model SEs are systematically far too low, about 50\% of their true values. In contrast, the SEs of the IL-HTE model are much better calibrated regardless of the level of IL-HTE in the data, an important result for accurate statistical inference.

The inflation of SEs occurs because IL-HTE accounts for an additional source of uncertainty in the estimation of the ATE. That is, if there is IL-HTE in the population of items from which a PROM is constructed, any finite draw of items for a PROM administration will have a sample ATE that varies from the population ATE due to sampling error. The random slope variance of the IL-HTE model ($\sigma^2_\zeta$) accounts for this source of uncertainty and provides adjusted estimates of the SE of the treatment effect, whereas the constant treatment effect model provides uncertainty consistent with the set of realized items treated as fixed \citep{gilbert2023modeling, miratrix2021applied}. The implication of this result is that researchers may be vastly overestimating their precision when they assume a constant effect model and intend to generalize their results to a latent trait that could have been measured with an alternative set of items.

\begin{figure}[tb]
    \centering
    \includegraphics[width=1\linewidth]{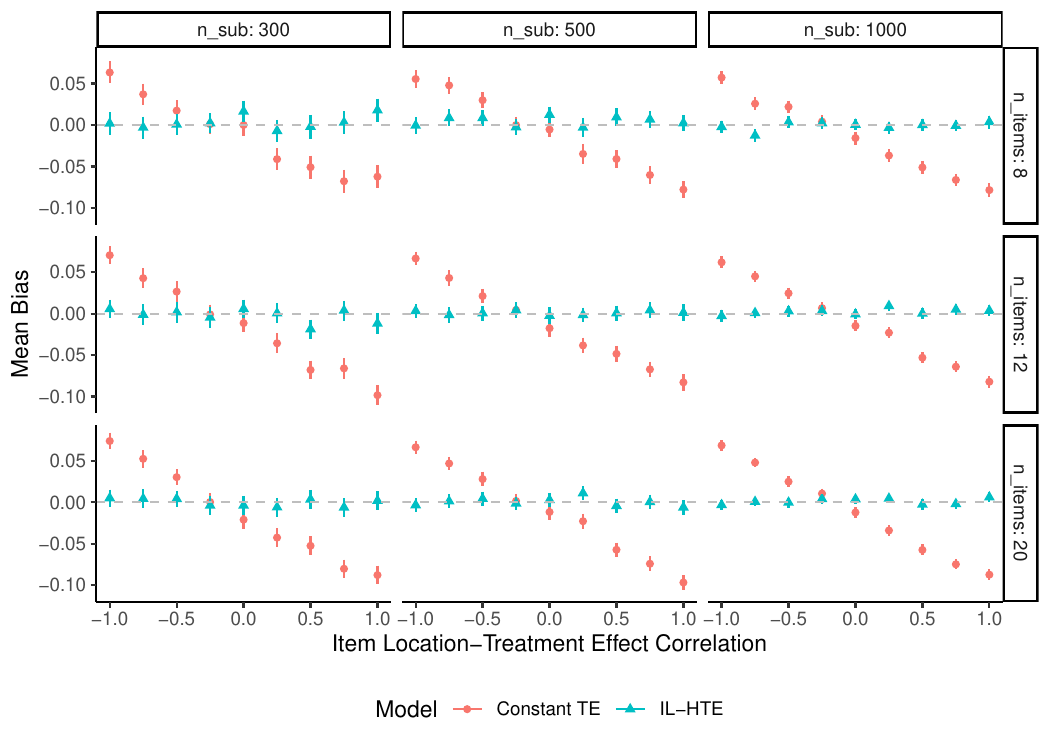}
    \caption{Estimated Bias in Treatment by Baseline Interaction Term}
            \justify \footnotesize 
    Notes: The y-axis shows the mean bias in the treatment by baseline covariate interaction term and the x-axis shows the item location by item-specific treatment effect correlation ($\rho$). \textit{n\_sub} is the number of subjects and \textit{n\_items} is the number of items. The bars show Monte Carlo 95\% CIs.
    \label{fig:rho_bias}
\end{figure}

Our second set of simulations examines the impact of $\rho$ on treatment by baseline covariate interactions also confirms previous simulation findings from dichotomous items \citep{gilbert2023disentangling}. That is, $\rho$ induces a spurious interaction between treatment and baseline variables. Figure \ref{fig:rho_bias} shows the bias in an estimated baseline by treatment interaction term (the true data-generating value is zero), and we see that high magnitudes of $\rho$, both positive and negative, induce a substantial bias in the estimated treatment by baseline interaction term in the constant treatment effect model. In contrast, the IL-HTE model that allows for $\rho$ successfully eliminates this bias. The pattern is identical across sample size and number of PROM items. This result suggests that treatment by baseline covariate interaction effects estimated on PROMs should be interpreted cautiously when item-level data is not available \citep{domingue2022ubiquitous, gilbert2023disentangling}.

\subsection{Empirical Application}

Model results are presented in Table \ref{table:regtable}. Models 1A and 1B present the conventional analysis of the standardized HDRS-17 sum scores as the outcome variables. We see that the average effect of SSRIs is negative and statistically significant, about .26 SDs ($p<.001$). A test for a treatment by baseline depression interaction is not significant in Model 1B. A conventional analysis focused on person-dependent HTE might stop here and conclude that the effects of SSRIs are homogeneous. Models 2A, 2B, and 2C allow for IL-HTE by directly modeling the item-level outcome data. Model 2A replicates the initial finding, showing a negative effect of SSRIs on depression, at about -.20 units on the logit scale ($p<.001$). Comparing the constant treatment effect Model 2A to the IL-HTE Model 2B, we see substantial IL-HTE in the data, with large increases to the log likelihood for the IL-HTE models and $\sigma_\zeta=.18$ for Model 2B, an SD nearly as large as the point estimate for the ATE itself. Consequently, we can see that $\text{SE}(\hat{\beta_1})$ almost doubles, capturing the added uncertainty of which items were selected for the HDRS-17 from the population of potential items that could have been selected. In other words, the estimated treatment effect of SSRIs on depression measured with a different set of items from those that could have been selected might be quite different than what was observed on these items. The approximate doubling of $\text{SE}(\hat{\beta_1})$ is equivalent to reducing the effective sample size by a factor of four.

\begin{landscape}
    \include{reg_table}
\end{landscape}

An intuitive way to interpret the IL-HTE parameter $\sigma_\zeta$ is as a measure of generalizability. That is, we can calculate a 95\% prediction interval (PI) for a range of item-specific treatment effects on \textit{out-of-sample} depression items (i.e., items that are similar to those on the HDRS-17), using the formula $\text{PI} = \hat\beta_1 \pm 1.96 \sqrt{\sigma^2_\zeta + Var(\hat\beta_1)}$ \citep[p. 130]{borenstein2009introduction}. Applied to Model 2B, we see that,

\begin{align}
    \text{PI}_{\text{M2B}} &= -.204 \pm 1.96 \sqrt{.034 + .003} \\
    &= -.204 \pm .377 \\
    &= [-.581,.173],
\end{align}

\noindent which suggests that treatment effects on items measuring other symptoms that could reasonably be included in a depression rating scale could show anywhere from large negative effects of SSRIs to moderate \textit{positive} effects (i.e., be made worse by SSRIs), a key finding for the generalizability of SSRI effects on PROMs.

\begin{figure}
    \centering
    \includegraphics[width=1\linewidth]{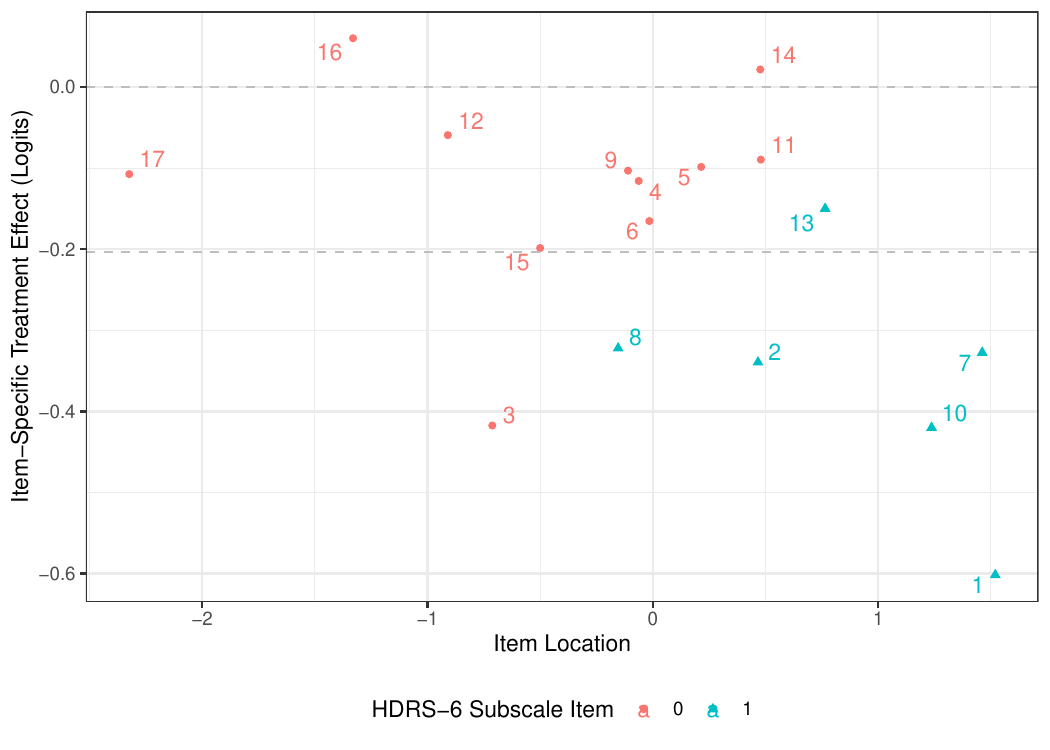}
    \caption{Correlation between Item Location and Item-Specific Treatment Effect Size}
         \justify \footnotesize 
    Notes: The y-axis shows Empirical Bayes estimates for item-specific treatment effect ($\beta_1+\zeta_i$) plotted against on Empirical Bayes estimates of item location in the control group ($b_i$) on the x-axis. The estimates are derived from Model 2B. A table of these values for each item is presented in our online supplement.
    \label{fig:blups_scatter}
\end{figure}

The correlation between item location and treatment effect size of about $\rho=-.5$  in Model 2B suggests that the most commonly endorsed symptoms saw the largest (negative) effects, as displayed in Figure \ref{fig:blups_scatter}, which plots empirical Bayes estimates of item-specific treatment effects ($\beta_1+\zeta_i$, y-axis) against item location in the control group ($b_i$, x-axis). As discussed earlier, $\rho$ can create causal identification challenges by inducing spurious interaction effects when not appropriately modeled, as shown in our simulation results in Figure \ref{fig:rho_bias}. We examine this phenomenon in our online supplement by fitting additional models that allow for a baseline depression by SSRI interaction, and see that the inclusion of $\rho$ in the model shifts the point estimate of the interaction effect, in line with our simulation results and prior research. However, the shift in the estimated interaction term in our data set is not large in magnitude, changing from -.015 in the constant effects model to -.024 in the IL-HTE model, and neither is statistically significant. Figure \ref{fig:blups_scatter} also highlights the HDRS-6 items, and it appears that the treatment effects on the HDRS-6 subscale are systematically larger than treatment effects on the 11 remaining items, and that the negative correlation may be partially or entirely driven by the larger effects on the HDRS-6 subscale.

\begin{figure}
    \centering
    \includegraphics[width=1\linewidth]{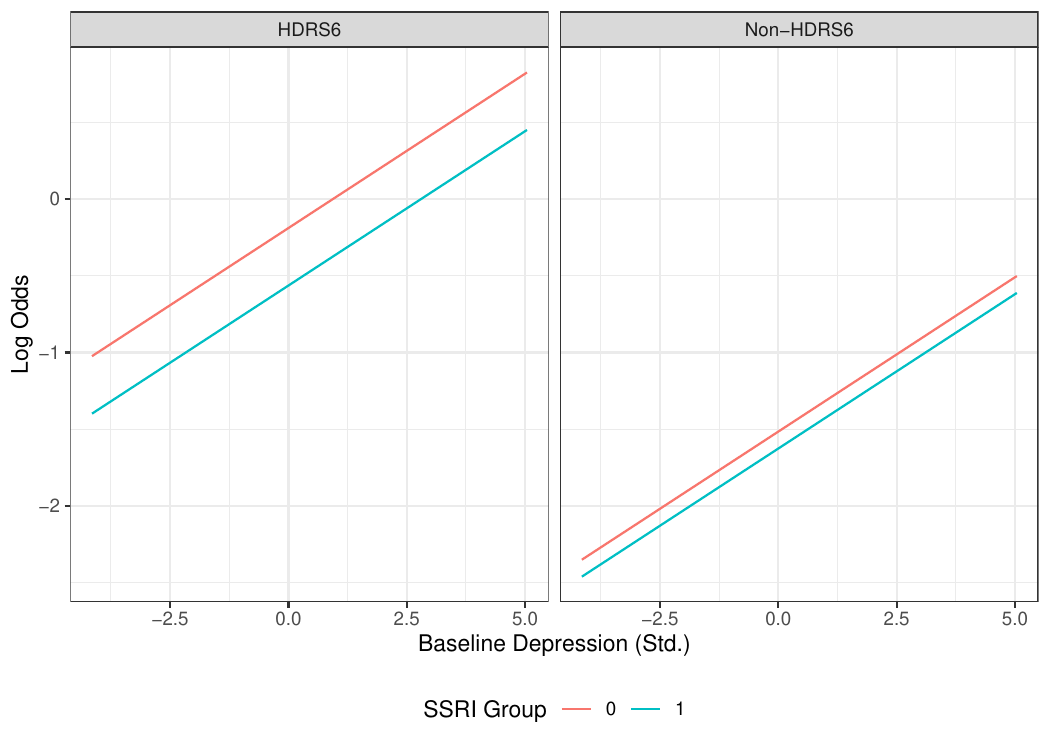}
    \caption{Conditional Average Treatment Effects of SSRIs on HDRS-6 Items and the Remaining 11 Items}
         \justify \footnotesize 
    Notes:  The panels show conditional average treatment effects for the HDRS-6 items and the non-HDRS-6 items. The y-axis shows the fitted log-odds of endorsing the average category (or higher) on the average item on each subscale, and the x-axis shows standardized HDRS-17 sum scores at baseline. The estimates are derived from Model 2C.
    \label{fig:ham6_logit}
\end{figure}

Model 2B shows substantial IL-HTE and make appropriate adjustments to our SEs, and as such is preferable to Models 2A. However, with such a large value of $\sigma_\zeta$, a natural question is what item characteristics might explain this IL-HTE. Accordingly, Model 2C includes HDRS-6 subscale by SSRI interaction effects to allow the treatment effect to systematically differ between the HDRS-6 items and the remaining 11 items \citep{gilbert2023modeling, gilbert2023tutorial, kim2023longitudinal}. Model 2C confirms that treatment effects on HDRS-6 items are systematically larger than the other 11 items ($\beta = -.265, p < .001$). The main effect of treatment on the remaining 11 items is negative and statistically significant, though much reduced in magnitude ($\beta=-.111, p<.05$). These results are in line with prior analysis of differential effects when each subscale is considered separately \citep{hieronymus2019influence}; one advantage of the IL-HTE model is that we obtain a direct hypothesis test of differences in effect size by subscale in a single model. The results of Model 2C are displayed in Figure \ref{fig:ham6_logit}, which shows the fitted log-odds of exceeding the average category on an average item in each subscale (y-axis) by baseline depression (x-axis) and treatment status (color). The CATE is represented by the vertical distance between the fitted lines, and we can see that it is much larger for the HDRS-6 items.

The larger effects on the HDRS-6 subscale have important policy implications for clinical trial evaluation and interpretation, because, ``if researchers could somehow \textit{a priori} select those items known to be more sensitive to the treatment, they would obtain a larger measured treatment impact as an artifact of the selected items, rather than a truly more effective treatment'' \citep[p. 895]{gilbert2023modeling}. In the context considered here, evaluations of SSRIs using HDRS-6 might appear to be more effective than those using the full HDRS-17, not because the treatment is more effective on depression as a whole, but because the specific symptoms of depression assessed on the HDRS-6 subscale are more sensitive to SSRIs. Conversely, it is also the case that some symptoms that are common in depression covary also with other psychiatric and somatic conditions, as well as with age and sex; insomnia being more common in old age, for example. In such cases, some ``depressive symptoms'' could be expected to persist after remission of the depressive episode because in such a case they are not causally related to depression, but to an alternative explanation (e.g., added exogeneous causes of the item responses above $\eta_{ij}$ in Figure \ref{fig:dag}). That is, such symptoms may reflect some construct-irrelevant variance \citep{downing2002threats}. Indeed, it is well known that symptomatic remission as measured by the HDRS-17 does not map particularly well to patient-defined remission (in either direction) \citep{zimmerman2012, zimmerman2012b}; such an inference is similarly supported by our IL-HTE analysis. 

We can calculate an approximate $R^2$ of the proportion of IL-HTE explained by the SSRI by HDRS-6 interaction effect comparing $\sigma^2_\zeta$ between models B and C \textbf{(i.e., Eqn. \ref{eq_r2})}. We see that the SSRI by HDRS-6 interaction explains 50\% percent of the IL-HTE, and that after accounting for the differential effects by subscale, our estimate of $\rho$ goes nearly to 0, in line with Figure \ref{fig:blups_scatter}, suggesting that the majority of the IL-HTE in the data is explained by the differential effects by subscale.

As a sensitivity check, we report the results of Model 2B for each of the 24 RCTs separately in our online supplement. We find that the results are stable across the disaggregated trials, with moderate negative effects of SSRIs on average and with the mean estimate of $\sigma_\zeta$ at .13 (IQR .08, .19), suggesting that our results are not driven by patient heterogeneity across the pooled sample.

\section{Discussion}\label{discussion}

Models for heterogeneous treatment effects (HTE) have typically focused on person characteristics as moderators of treatment efficacy. While valuable, such approaches can generate false positives given the inference, generalizability, and identification challenges that arise when the outcome of interest is a latent variable such as a PROM constructed from a set of items. In this study, we extend novel Item Response Theory methods to the estimation of item-level HTE (IL-HTE) using an analysis of clinical trial data of the effects of SSRIs on depression as measured by the 17-item Hamilton Depression Rating Scale as an illustration. As such, our study represents an important contribution to epidemiological methodology, from both statistical and substantive perspectives.

When IL-HTE is present in the data but ignored in the model, the standard errors of treatment effects are underestimated, leading to spurious estimates of precision and potential false positives. For example, the near doubling of the SE of the treatment effect we observed in our empirical analysis is equivalent to a reduction in the sample size by a factor of four, a finding with implications for both statistical inference generally and prospective power analysis. The IL-HTE model better accounts for the uncertainty of which items were chosen for the PROM and provides a metric for how generalizable our results might be if we used different items to assess treatment impact. Similarly, the ability to model subscale effects can provide insight into which symptom groups are most sensitive to treatment, allowing researchers to test explicit hypotheses about differential subscale effects. Furthermore, correlation between item location and item-specific treatment effect sizes can induce spurious interactions between treatment and baseline characteristics that only item-level analysis can appropriately identify. While spurious interactions did not emerge in our empirical illustration, it would be prudent to complement analyses that examine HTE by baseline covariates by the kind of item-level analysis advocated here to ensure accurate identification of interaction effects. Even when HTE by baseline covariates is not of primary interest, non-zero correlations will also create bias in the \textit{main} effects of covariates, because the main effect in a model without interaction is an average of the effects in each subgroup, weighted by subgroup sample size.

Substantively, the application of the IL-HTE model to the placebo-controlled SSRI trials showed substantial IL-HTE in the HDRS-17. Estimates of the standard deviation of item-specific treatment effects were nearly as large as the point estimates for the average treatment effect themselves, and standard errors nearly doubled when we accounted for IL-HTE in the model. Prediction intervals for out-of-sample items suggested that the impact of SSRIs could be anywhere from slightly harmful to strongly beneficial. Such a fine-grained level of insight on the extent of the impact of SSRIs on depression would be masked by a traditional analysis of a single-number summary outcome, such as a sum score.

Furthermore, our analysis of subscale effects on the HDRS-6 items showed systematically higher treatment effects on HDRS-6 than the other eleven items, a finding that has potentially critical policy implications for how we judge the effectiveness of treatments using different outcome measures for the same underlying construct such as depression. The HDRS-6, which was designed to be a unidimensional measure of depressive severity, was developed well before modern SSRIs came to market and it is therefore unlikely that researchers developing the HDRS-6 would have chanced upon a collection of items that would make SSRIs appear particularly effective in the future \citep{bech1975,bech1981hamilton}, especially considering that all efforts aimed at developing unidimensional subscales have reached subscales that closely resemble the HDRS-6 (i.e., always including the items depressed mood, feelings of guilt, work and activities, and psychic anxiety) \citep{bagby2004}. Nevertheless, the HTE by subscale has important implications for the interpretation of HDRS-6 or other subscales as outcome measures given that it demonstrates the extent to which efficacy estimates are conditional on which rating instrument is applied \citep{ruhe2005clinical, luckenbaugh2015rating}. 

In the same vein, SSRIs---the antidepressant class examined in this study---have well-known side effects related to sexual dysfunction, decreased appetite, gastrointestinal complaints, and insomnia \citep{ferguson2001}. These side effects, which are also present in healthy volunteers \citep{knorr2019}, have been shown to correlate with the scores of the corresponding HDRS items, introducing another way in which ratings of individual depressive symptoms may diverge from the latent variable they are intended to measure \citep{hieronymus2021}. In particular, other antidepressants, such as amitriptyline and mirtazapine have the opposite effects in that they are hypnotics and increase appetite and could therefore potentially introduce the opposite biases, that is, making them appear as more effective antidepressants when measured on a scale including many such items.

\subsection{Limitations}

Clearly, IL-HTE has both statistical and substantive implications for the analysis of treatment effects in clinical trials and epidemiology, but we acknowledge the following four limitations of the IL-HTE approach, based on the limitations described in previous research \citep{gilbert2023modeling, gilbert2023disentangling}. First, it is unknown how common IL-HTE is in clinical trial or epidemiological PROMs data, and we leave this largely as an open question. Some systematic evidence comes from education research, where an analysis of 15 RCTs showed substantial IL-HTE, including a case where 40\% of the overall treatment effect was driven by a single test item \citep{ahmed2023heterogeneity}. Therefore, future research using the IL-HTE model in other clinical trial or epidemiological contexts has the potential to shed light on how widespread the IL-HTE phenomenon is in this field.

Second, estimating the IL-HTE model requires the availability of item-level data, which may not be available, particularly in secondary analyses. As such, a practical implication of our results is that researchers should heed calls to share item-level outcome data \citep{domingue2023item}, so that researchers and secondary analysts can evaluate the extent of IL-HTE in a data set given the statistical and substantive insights allowed by IL-HTE analysis.

Third, the use of latent variable models such as the explanatory item response model may be more difficult to interpret and justify to practitioners, particularly when the coefficients are on the logit scale \citep{breen2018interpreting, mood2010logistic}. One approach to improve communicability is to convert the logit coefficients to standardized effect sizes by dividing the regression coefficient by the estimated standard deviation of the latent variable $\theta_j$, a value that can be derived from the results of the model \citep[pp. 907-908]{gilbert2023modeling}.

Finally, the explanatory item response model can be substantially more computationally demanding than alternative models such as OLS regression because the cross-classified multilevel structure of the item response data requires numerical integration approaches \citep{rabe2022multilevel} or Markov chain Monte Carlo (MCMC) methods in Bayesian applications \citep{burkner2021brms, gilbert2023tutorial}. Thus, when the number of items and persons is large, IL-HTE methods may become computationally prohibitive.

\subsection{Conclusion}

In sum, applying measurement and psychometric principles to causal inference in epidemiological and public health research provides a powerful opportunity anywhere multi-item patient-reported outcome measures are used to assess treatment impact. By using the IL-HTE model, both clinical and epidemiological researchers can obtain more accurate statistical inference, better estimates of generalizability to new symptoms, and unbiased estimates of interaction effects, all essential qualities for clinical trial and epidemiological research that aims to inform public health policy.

\newpage

\section{Declarations}

\textbf{Research Ethics}

Not applicable.

\textbf{Informed Consent}

Not applicable.

\textbf{Author Contributions}

Conceptualization: Author 1, Author 4

Methodology: Author 1, Author 4

Software: Author 1, Author 2

Formal Analysis: Author 1

Writing---original draft preparation: Author 1

Writing---review and editing: Author 1, Author 2, Author 3, Author 4

\textbf{Competing Interests}

Authors 1 and 4 report no conflicts of interest.

Author 2 has received speaker's fees from Janssen Pharmaceuticals in the last five years and is a board member of the Swedish Serotonin Society. 

Author 3 has received speaker's fees from Janssen Pharmaceuticals in the last five years.

\textbf{Research Funding}

This work was funded in part by the Jacobs Foundation.

\textbf{Data Availability}

The data analyzed in this study are proprietary. However, we share a simulated version of the data set based on the results of our models in our supplement. The code for this article is available at the following URL: \url{https://researchbox.org/2494}.

The full HDRS-17 instrument is available at the following URL: \url{https://dcf.psychiatry.ufl.edu/files/2011/05/HAMILTON-DEPRESSION.pdf}

\newpage

\printbibliography

\end{document}

%% file: reg_table.tex
\begin{table}
\begin{center}
\begin{tabular}{l c c c c c}
\hline
 & M1A & M1B & M2A & M2B & M2C \\
\hline
Intercept                    & $.183^{***}$  & $.183^{***}$  & $-.404$        & $-.417$        & $-.881^{***}$  \\
                             & $(.024)$      & $(.024)$      & $(.227)$       & $(.241)$       & $(.225)$       \\
1 = SSRI                     & $-.262^{***}$ & $-.262^{***}$ & $-.224^{***}$  & $-.204^{***}$  & $-.111^{*}$    \\
                             & $(.029)$      & $(.029)$      & $(.027)$       & $(.052)$       & $(.050)$       \\
Baseline Depression (Std.)   & $.229^{***}$  & $.254^{***}$  & $.200^{***}$   & $.201^{***}$   & $.201^{***}$   \\
                             & $(.013)$      & $(.024)$      & $(.012)$       & $(.012)$       & $(.012)$       \\
SSRI x Baseline Depression   &               & $-.037$       &                &                &                \\
                             &               & $(.029)$      &                &                &                \\
1 = HDRS6 Item               &               &               &                &                & $1.328^{***}$  \\
                             &               &               &                &                & $(.376)$       \\
SSRI x HDRS6                 &               &               &                &                & $-.265^{***}$  \\
                             &               &               &                &                & $(.072)$       \\
Threshold 2                  &               &               & $-.764^{***}$  & $-.772^{***}$  & $-.772^{***}$  \\
                             &               &               & $(.016)$       & $(.016)$       & $(.016)$       \\
Threshold 3                  &               &               & $-2.618^{***}$ & $-2.642^{***}$ & $-2.642^{***}$ \\
                             &               &               & $(.029)$       & $(.029)$       & $(.029)$       \\
Threshold 4                  &               &               & $-3.818^{***}$ & $-3.870^{***}$ & $-3.872^{***}$ \\
                             &               &               & $(.075)$       & $(.076)$       & $(.076)$       \\
\hline
Num. obs.                    & $5313$        & $5313$        & $132326$       & $132326$       & $132326$       \\
AIC                          & $$            & $$            & $146967.819$   & $146838.173$   & $146827.259$   \\
BIC                          & $$            & $$            & $147046.163$   & $146936.103$   & $146944.775$   \\
Log Likelihood               & $$            & $$            & $-73475.910$   & $-73409.086$   & $-73401.629$   \\
Num. groups: PID             & $$            & $$            & $5313$         & $5313$         & $5313$         \\
Num. groups: itemID          & $$            & $$            & $17$           & $17$           & $17$           \\
Var: PID (Intercept)         & $$            & $$            & $.562$         & $.571$         & $.570$         \\
Var: itemID (Intercept)      & $$            & $$            & $.867$         & $.985$         & $.549$         \\
Var: itemID SSRI             & $$            & $$            & $$             & $.034$         & $.017$         \\
Cov: itemID (Intercept) SSRI & $$            & $$            & $$             & $-.090$        & $-.004$        \\
\hline
\multicolumn{6}{l}{\scriptsize{$^{***}p<0.001$; $^{**}p<0.01$; $^{*}p<0.05$}}
\end{tabular}
\caption{Explanatory item response models fit to the SSRI Data}
\label{table:regtable}
\end{center}

        \justify \footnotesize 
    Notes: PID = Person identifier. itemID = item identifier. Models 1A and 1B are OLS models of the standardized HDRS-17 sum score. Models 2A, 2B, and 2C are Rating Scale Models that assume a constant treatment effect, randomly varying IL-HTE, and treatment by subscale interactions, respectively.

\end{table}